\documentclass[aps,prd,preprint,tightenlines,floatfix,showpacs,
groupedaddress]{revtex4}

\usepackage{subfigure}
\usepackage{graphicx}

\begin{document}

\vspace{1cm}

\date{March 27, 2007}

\title{WIMP annihilation in caustics}

\author{Aravind Natarajan}

\affiliation{Institute for Fundamental Theory\\
   Department of Physics\\
   University of Florida\\
   Gainesville, FL,  32611-8440\\}

\begin{abstract} 
 The continuous infall of dark matter with low velocity dispersion from all directions in a galactic halo leads to the formation of caustics which are very small scale ($\sim$parsec) high density structures. If the dark matter is made up of SUSY neutralinos, the annihilation of these particles produces a characteristic spectrum of gamma rays which in principle could be detected. The annihilation signal at different energy bands is computed and compared with the expected gamma ray background. 

\vspace{0.34cm}
\noindent
\end{abstract}
\maketitle

\section{Introduction\label{one}}

According to the standard theory of cosmological structure formation, most of the mass in galactic halos is composed of non-baryonic matter with negligible primordial velocity dispersion, commonly called cold dark matter (CDM), the constitution of which remains a mystery.  Many theories have been proposed to account for the dark matter.  Dark matter candidates may be non-thermal relics or thermal relics, depending upon the underlying particle physics. Non-thermal relics are particles which decoupled from the standard model particles without being in thermal equilibrium with them, the best example being the axion~\cite{ref1}. Thermal relics on the other hand, were once in thermal equilibrium with other particles, but decoupled when their interaction rate became comparable to the Hubble rate. These particles are collectively known as Weakly Interacting Massive Particles or WIMPs~\cite{ref2}.
 
There are many collaborations currently trying to detect dark matter, which include ADMX~\cite{ref3}, DAMA/NaI~\cite{ref4}, DAMA/LIBRA~\cite{ref5}, CDMS~\cite{ref6}, XENON~\cite{ref7}, EDELWEISS~\cite{ref8}, ZEPLIN~\cite{ref9}, etc.  Dark matter detection experiments are usually of two kinds - direct detection experiments look for the recoil that occurs when a WIMP scatters off a target nucleus, while indirect detection experiments look for standard model particles which result from dark matter particle annihilation. In this article, we investigate the flux of gamma ray photons produced by WIMP annihilation in dark matter caustics. Previous work on particle annihilation in  caustics includes \cite{ref10,ref11,ref12,ref13,ms}.  For the possibility of detecting caustics by gravitational lensing, see \cite{ref14,ref15,onemli}. Caustics and their associated cold flows are also relevant to direct detection experiments \cite{ref16,ref17,ref18,ref19,ref20}.
 
Cold dark matter particles exist on a thin 3-dimensional hypersurface in phase space. To obtain the density in physical space, we must make a mapping from phase space to physical space. Caustics are locations where this mapping is singular~\cite{asz,ref21, ref22, ref23, ref24, ref25, ref26, ref27, ref28, ref29, ref30, ref31, ref32,zs}. Caustics are made up of sections of the elementary catastrophes, which are structurally stable and have a definite geometry that depends on the structure of the phase space manifold. In the limit of zero velocity dispersion, caustics are singularities in physical space. In real galactic halos, the velocity dispersion cuts off the divergence \cite{zs}. 

It is important to ask whether the finite velocity dispersion of dark matter can soften the density contrast of caustics to the extent that the caustics are made irrelevant. The inner regions of phase space may well have thermalized over the age of the Universe. The infall of thermal dark matter does not produce observable caustics. However, it is believed that in large galaxies like our own, there is a lot of dark matter well outside the virial radius, yet gravitationally bound to the halo. These dark matter particles are in the process of thermalizing and have velocity dispersions that are small compared to the virial velocity dispersion of the galaxy~\cite{ref28,ref30}. The dark matter particles reaching us today for the first time for example, are cold and are likely to produce a physically significant caustic. Similarly the particles that have fallen into the inner regions of the halo only a few times in the past may be expected to produce observable caustics. Thus caustics are formed if the outer regions of phase space are well resolved. An analysis of the rotation curves of 32 galaxies~\cite{ref33} seems to provide evidence for the existence of dark matter caustics. The rotation curve of our galaxy also shows rises at the expected locations of the caustics~\cite{ref34}.

Perhaps the best example of caustics on galactic scales is the occurrence of shells around giant elliptical galaxies~\cite{ref35, ref36, ref37} (e.g. NGC 3923, see~\cite{ref37}). These shells are caustics in the distribution of starlight. They form when a dwarf galaxy falls into the gravitational potential of a much larger galaxy and is assimilated by it. The velocity dispersions of the stars of the dwarf galaxy are much smaller than the virial velocity dispersion of the giant galaxy. The infall of stars is therefore cold and presumably, collisionless. The occurrence of these caustics show us that cold flows are not unstable and that caustics do not require special initial conditions to form~\cite{ref30}. This leads us to believe that the infall of non-thermal dark matter also produces caustics.

There are two kinds of caustics - outer and inner~\cite{ref29}. The outer caustics are thin topological spheres surrounding galaxies (like the caustic of stars discussed previously). They are simple fold catastrophes and typically occur on scales of 100's of kpc for a galaxy like our own. The inner caustics have a more complicated geometry and are made up of sections of the higher order catastrophes~\cite{ref31}. Inner caustics typically occur on scales of 10's of kpc.

\begin{figure}[]
\includegraphics[width=3in,height=3in] {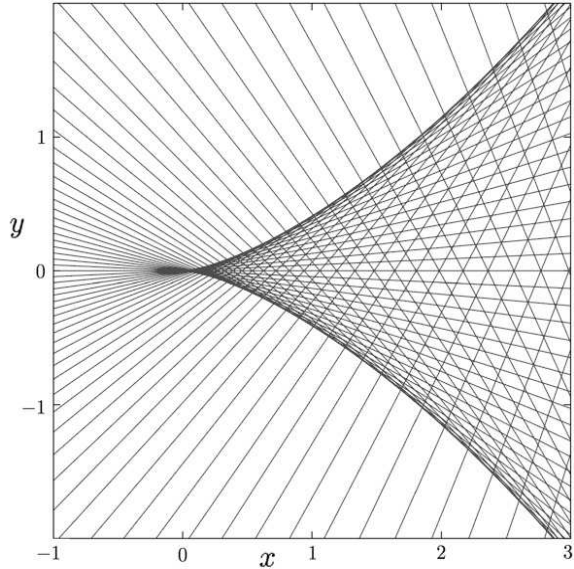}
\caption{ Dark matter trajectories forming a caustic. The caustic is the envelope of the family of dark matter trajectories. Thus the dark matter density is large at the location of the caustic. }
\label{fig1}
\end{figure}

To see the formation of inner caustics, let us consider the infall of a perfectly cold flow of dark matter. If the infall is exactly spherically symmetric, the particle trajectories are radial and the infall produces a singularity at the center. If instead, the dark matter particles possess some distribution of angular momentum with respect to the halo center, the particle trajectories are non-radial, particularly in the inner regions of the halo. Fig.~\ref{fig1} shows an example of non-radial infall (in cross section) with the result that the dark matter density is enhanced along two thin fold catastrophe lines (surfaces in 3-dim. space) which meet at a cusp catastrophe point (line in 3-dim. space)~\cite{ref38,ref39}. The caustic divides the $xy$ space into two regions - one region with 1 particle trajectory passing through each point (outer region) and the other with 3 particle trajectories passing through each point (inner region).

Close to the cusp, the dark matter trajectories are specified by the family of curves
\begin{equation}
\frac{y}{y_0} = \alpha^3 - \alpha \frac{x}{x_0} \\
\label{eq_1} 
\end{equation}
where $(x=0,y=0)$ are the co-ordinates of the cusp and $x_0,y_0$ are constants. $\alpha$ parametrizes the family. The caustic is the envelope of the family of curves, i.e. the locus of points tangent to the curves. The envelope is obtained by solving Eq.~\ref{eq_1} together with the tangency condition $3\alpha^2 - \frac{x}{x_0} = 0$, using which we obtain the caustic curve
\begin{equation}
\frac{y}{y_0} = \pm \frac{2}{3\sqrt{3}} \left( \frac{x}{x_0} \right )^{3/2}.
\label{eq_3}
\end{equation}
The density is proportional to the sum $ \sum_{i}   \left| 3 {\alpha_i}^2 - \left(x/x_0\right ) \right | ^{-1} $ where $\alpha_i$ are the real roots of the cubic Eq.~\ref{eq_1}. When $y=0$, the dark matter density is given by  $ \rho(x,0) = \rho_c  \left|\frac{x_0}{x}\right| $ for $x < 0$ and $ \rho(x,0) =  \rho_c \left( \frac{x_0}{x} + \frac{x_0}{2x} + \frac{x_0}{2x} \right ) = 2\,\rho_c \frac{x_0}{x}$ for $x > 0$ where $\rho_c$ is a constant. For $x=0$, we have $\rho(0,y) = \frac{\rho_c}{3} \left(\frac{y_0}{y}\right)^{2/3}$. Close to one of the fold lines and in the inner region, the density falls off as the inverse square root of the distance to the fold along the direction perpendicular to the fold.  For arbitrary $(x,y)$, we need to solve Eq.~\ref{eq_1} to obtain $\alpha_i$.  The general solution is worked out in~\cite{ref14}.

For arbitrary initial conditions, other catastrophes can occur. In general, we may parametrize the flow of cold dark matter particles by a three parameter label $\vec\alpha(\alpha_1, \alpha_2, \alpha_3)$ where $\alpha_1, \alpha_2, \alpha_3$ are three conveniently chosen parameters that describe the flow~\cite{ref29}. Let $\vec x(t, \vec \alpha)$ be the physical space location at time $t$, of the particle labeled $\vec\alpha$ and is obtained by solving the equations of motion for given initial conditions. The physical space density is proportional to the sum $ \sum_i \, | \partial \vec x / \partial \vec \alpha_i |^{-1} $ ( the Jacobian determinant of the map from $\vec\alpha$ space to $\vec x$ space ) and is infinite (for a perfectly cold flow) at those points where $ \partial \vec x / \partial \vec \alpha_i  = 0$. For a perfectly cold flow, it can be shown that such a singularity necessarily occurs~\cite{ref31}.

A perfectly cold flow cannot be realized in practice. The primordial velocity dispersion of WIMPs is estimated to be $ \delta v = 3 \times 10^{-7} \sqrt{ 100 \, \textrm{GeV} / m_\chi }$ km/s~\cite{ref29}. This causes the caustic surface to spread by an amount $ \delta a \approx \delta v \times a / v $~\cite{ref40} where $a$ is the outer turnaround radius (few hundred kpc) of the flow of particles forming the caustic and $v$ is the speed of the particles at the location of the (inner) caustic (few $\times$ 100 $\,$ km/s~\cite{ref19}). This minimum spread in the caustic surface sets an upper limit to the density. For the nearby caustics, we can estimate the spread in the caustic location
\begin{eqnarray}
\delta a &\sim & \frac{200 \times 3 \times 10^{-7}}{500} \, \textrm{kpc} \, \sqrt{\frac{100 \, \textrm{GeV}}{m_\chi} } \;  \nonumber \\
  &\sim& 10^{-4} \; \textrm{pc} \; \sqrt{\frac{100 \, \textrm{GeV}}{m_\chi} } 
\end{eqnarray}

\begin{figure}[!h]

\subfigure[]{\label{fig2-a}\includegraphics[width=3.2in,height=2in]
{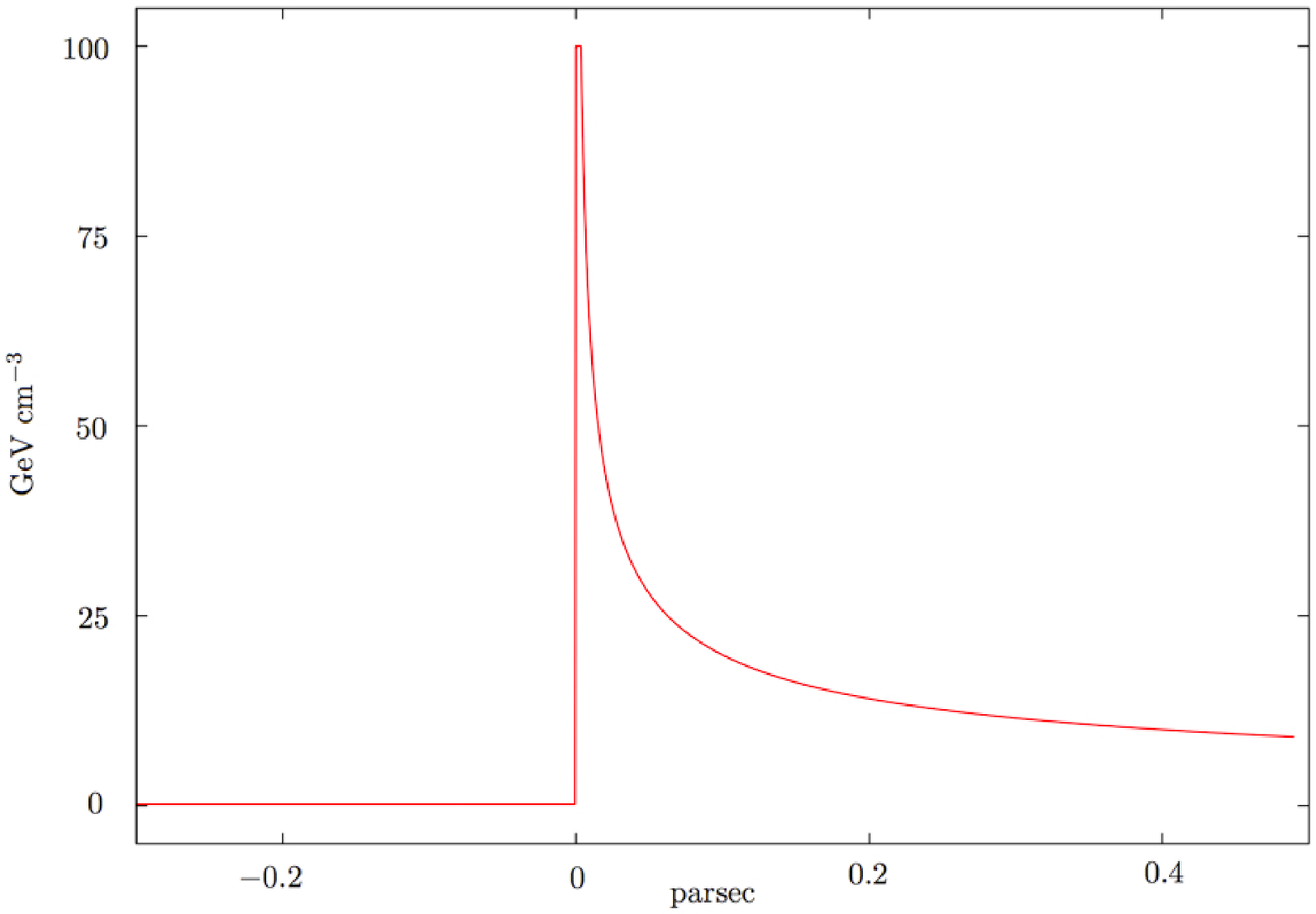}}
\subfigure[]
{\label{fig2-b}\includegraphics[width=3.2in,height=2in]
{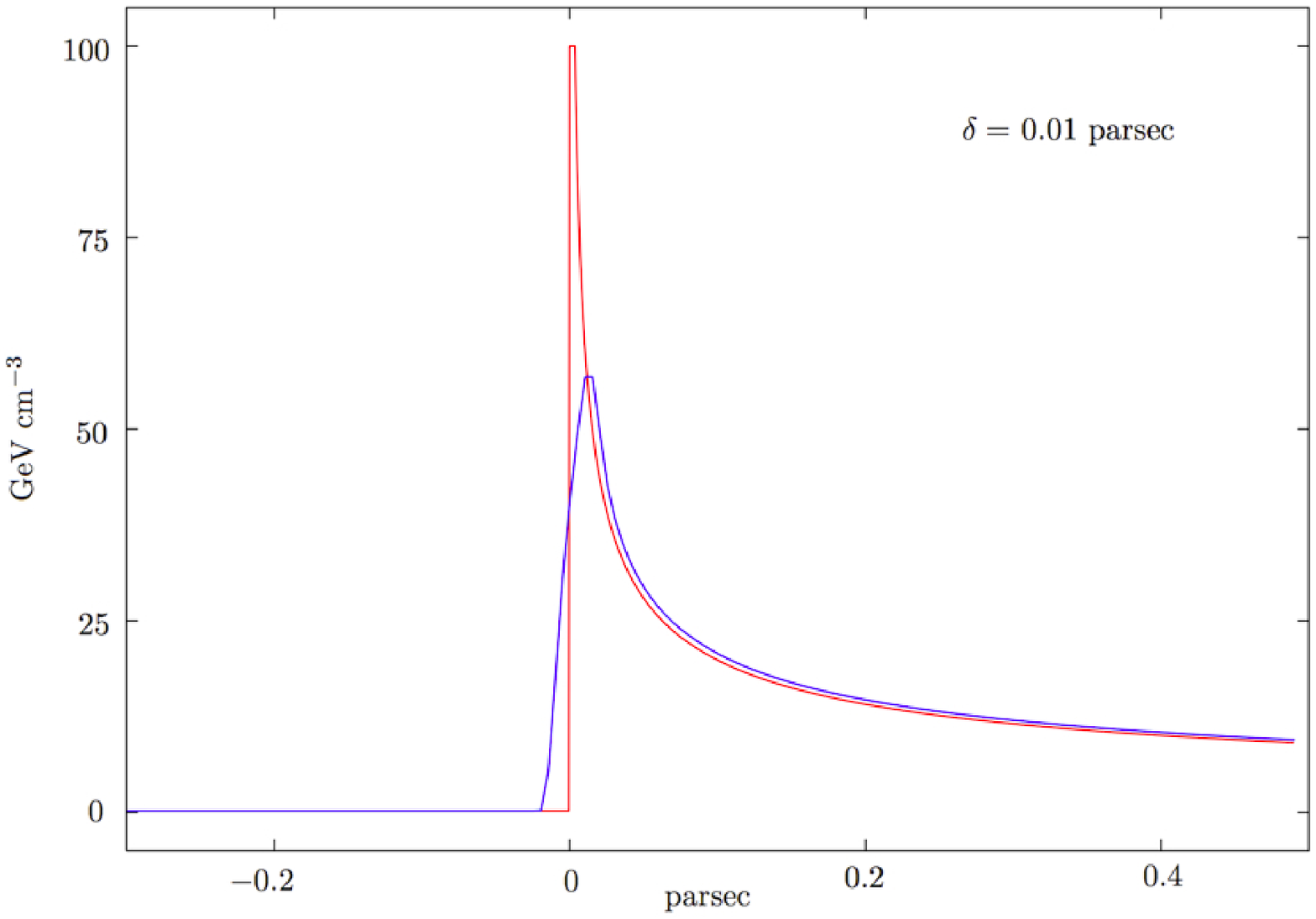}}
\subfigure[]
{\label{fig2-c}\includegraphics[width=3.2in,height=2in]
{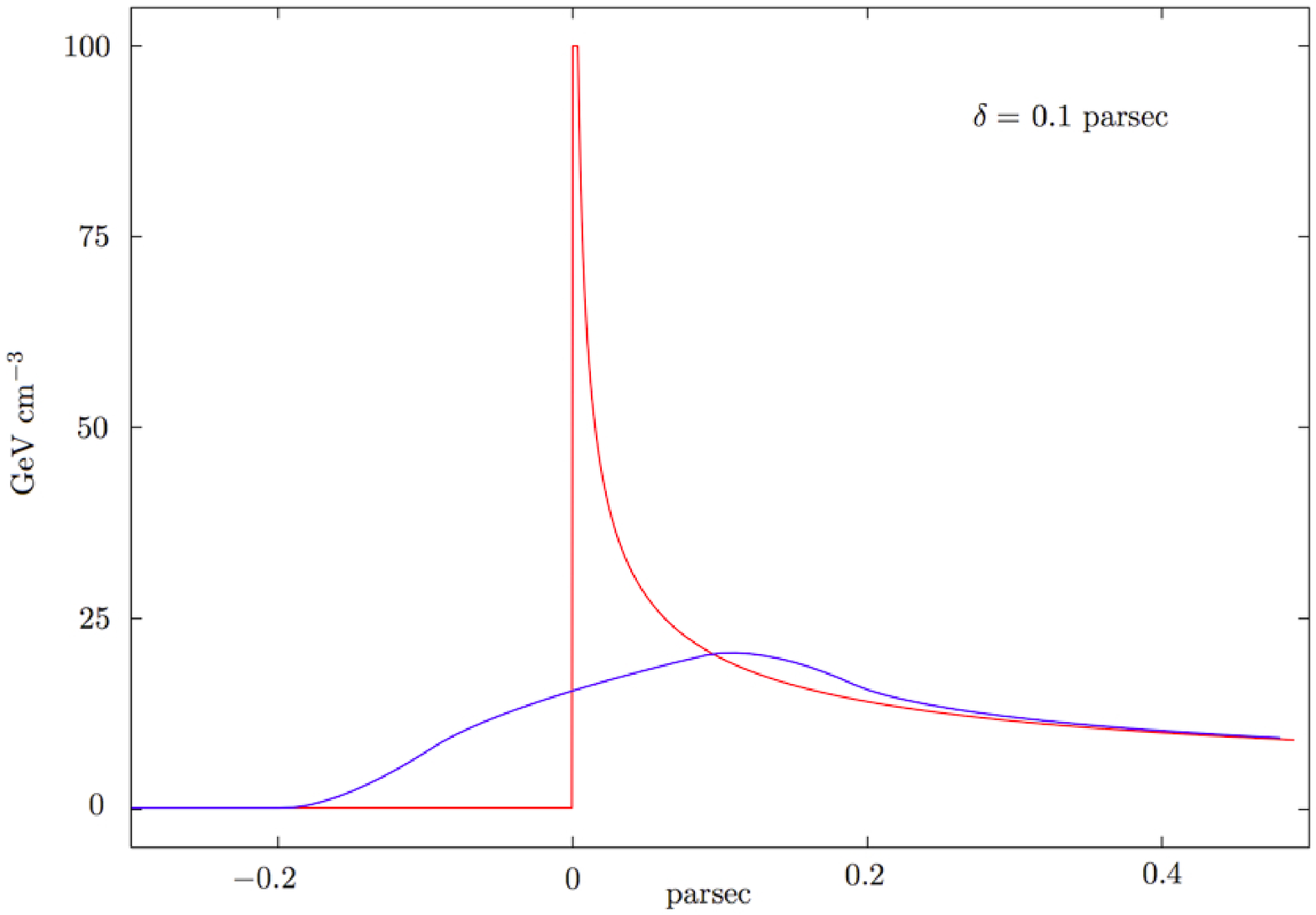}}
\subfigure[]
{\label{fig2-d}\includegraphics[width=3.2in,height=2in]
{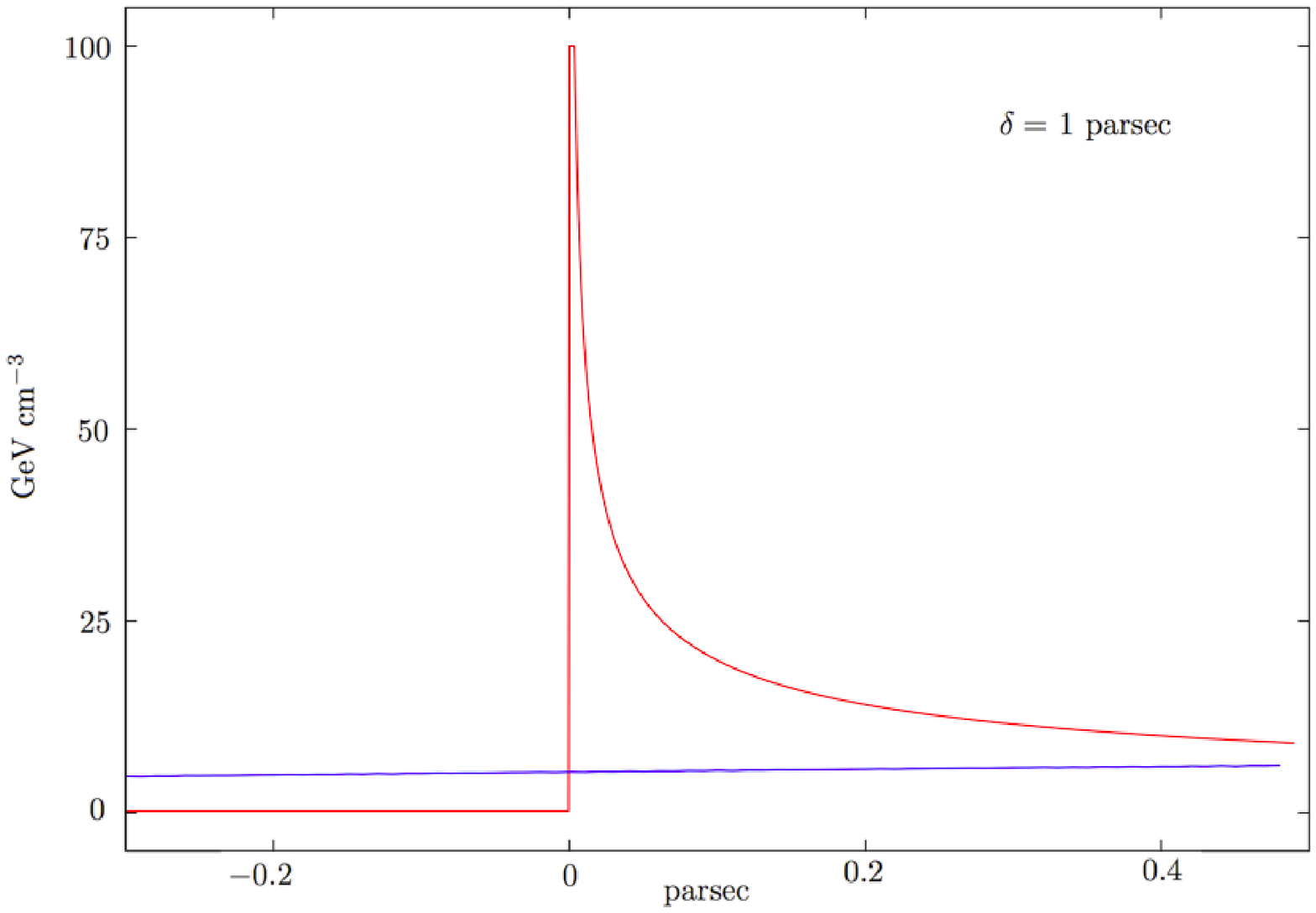}}

\caption{ (color online) The effect of averaging the density over a finite volume. Fig.~\ref{fig2-a} shows the variation of dark matter density (in Gev/cm$^3$) with distance to the caustic ( in parsec ). A cut-off density of 100 Gev/cm$^3$ was assumed. Figures~\ref{fig2-b},~\ref{fig2-c} and~\ref{fig2-d} show the effect of averaging the density over a cube of side $\delta$ for $\delta = 0.01, 0.1$ and $1$ parsec respectively. 
\label{fig2}}
\end{figure}

Let us now examine the effect of averaging the density over a finite region of space. Fig~\ref{fig2-a} shows the density along the line of sight, close to one of the fold lines of Fig~\ref{fig1}. Close to the fold, but outside it, the dark matter density is very small. The density changes abruptly at the location of the fold and becomes very large (100 GeV/cm$^3$ was chosen as the density cut-off), decreasing smoothly thereafter. Figures~\ref{fig2-b},~\ref{fig2-c} and~\ref{fig2-d} show the density averaged over a cube of side 0.01 parsec, 0.1 parsec and 1 parsec respectively. In Fig.~\ref{fig2-b}, we see that the averaged density follows the true density faithfully, though the maximum averaged density does not quite reach the cut-off value. Fig.~\ref{fig2-c} no longer shows a sharp rise in density and  Fig.~\ref{fig2-d} misses the caustic completely. We conclude that caustics are sub-parsec scale structures and are therefore difficult to resolve with large scale cosmological simulations which typically have spatial resolutions of 100's of pc.

\section{The annihilation flux}

In the minimal supersymmetric extension of the standard model (MSSM), a good candidate for the WIMP is the lightest neutralino which is a linear combination of the supersymmetric partners of the neutral electroweak gauge bosons and the neutral Higgs bosons. The characteristics of the annihilation signal depend both on the composition of the WIMP and its mass $m_\chi$. The line emission signal ( $\chi\chi \rightarrow \gamma\gamma, \chi\chi \rightarrow Z \gamma$ ) is loop suppressed and is therefore smaller than the continuum signal. The continuum flux (number of photons received with energies ranging from $E_1$ to $E_2$ per unit detector area, per unit solid angle, per unit time) is given by~\cite{ref12, ref41,ref42}
\begin{equation}
\Phi(E_1, E_2, \theta, \phi, \Delta\Omega) = S(E_1,E_2) \times \frac{< EM >}{4\pi} (\theta,\phi,\Delta\Omega)
\end{equation}
where
\begin{eqnarray}
S(E_1,E_2) &=& \int_{E_1}^{E_2} dE \, \sum_i b_i \, \frac{dN_{\gamma,i}}{dE}(E_\gamma) \; \frac{ <\sigma v> }{2\, m^2_\chi}  \nonumber \\
< EM >(\theta,\phi,\Delta\Omega) &=& \frac{1}{\Delta\Omega} \int_{\Delta\Omega} d\Omega \;\; EM(\theta,\phi) 
\end{eqnarray}
$dN_{\gamma,i}/dE$ is the number of photons produced per annihilation channel per unit energy, $b_i$ is the branching fraction of channel $i$ and $<\sigma v>$ is the thermally averaged cross section times the relative velocity. The factor of 2 in the denominator accounts for the fact that two WIMPs disappear per annihilation.   The quantity $EM(\theta,\phi)$ is called the emission measure and is the dark matter density squared, integrated along the line of sight, i.e.
\begin{equation}
EM(\theta,\phi) = \int_{los} dx \, \rho^2(x). 
\label{eq_7}
\end{equation}
and $< EM >(\theta,\phi,\Delta\Omega)$ represents the emission measure in the direction $(\theta,\phi)$ averaged over a cone of angular extent $\Delta\Omega$. We note that $S$ depends solely on the particle physics while $< EM >$ depends solely on the dark matter distribution.

\subsection{Estimating S}

Assuming that all the dark matter is composed of neutralinos, the quantity $ < \sigma v > $ is constrained by the known dark matter abundance~\cite{ref2}
\begin{eqnarray}
< \sigma v > &\approx & \frac{ 3 \times 10^{-27} \textrm{ cm$^3$ s $^{-1}$ } }{ \Omega_\chi h^2 }  \nonumber \\
& \approx & 3\times 10^{-26} \textrm{ cm$^3$ s $^{-1}$ } 
\end{eqnarray}
The quantity $dN_{\gamma}(E_\gamma) / dx$ for the dominant channels may be approximated by the form~\cite{ref43, ref44, ref45} $dN_{\gamma} / dx = a \, e^{-bx} / x$ where $x$ is the dimensionless quantity $E_\gamma / m_\chi$ and $(a,b)$ are constants for a given annihilation channel. The values of $(a,b)$ for the important channels are given in~\cite{ref43}. Using these values, we may calculate the number of photons produced per annihilation within a specified energy range. Let us consider four energy bands: Energy Band I with photon energies from 30 MeV to 100 MeV, Band II with energies from 100 MeV to 1 GeV, Band III with energies from 1 GeV to 10 GeV and Band IV containing photon energies 10 GeV upto $m_\chi$. The values of $N_\gamma / m^2_\chi$ are tabulated for the different  energy bands, for $m_\chi = 50, 100, 200$ GeV.

\begin{table}[!h]
\caption{ $ N_\gamma / {m^2_\chi} $ in units of $10^{-4} $ GeV$^{-2}$ for $m_\chi = 50 \, \textrm{GeV} $ } 
\centering 
\begin{tabular}{l c c c c } 
\hline\hline 
Channel & I  & II & III &  IV \\ [0.5ex] 
\hline 
    $ WW,ZZ (0.73, 7.76)$   &  106.8  &   85.2   &  18   &  0.52  \\
    $b \bar b  (1, 10.7)$     &  146     &  114.4  &  21.6  & 0.32 \\
    $ t \bar t (1.1, 15.1) $   &  160  &   122.4  &   19.6  &   0.12  \\
    $ u \bar u (0.95, 6.5)$  &  139.2  & 111.6  &   25.2  &   0.96 \\ [1ex] 
\hline 
\end{tabular}
\label{table:table1} 
\end{table}

\begin{table}[!h]
\caption{ $ N_\gamma / {m^2_\chi} $ in units of $10^{-4} $ GeV$^{-2}$ for $m_\chi = 100 \, \textrm{GeV} $ } 
\centering 
\begin{tabular}{l c c c c } 
\hline\hline 
Channel & I  & II & III &  IV \\ [0.5ex] 
\hline 
    $ WW,ZZ (0.73, 7.76)$   &  38  &   30.8   &  7.9   &  0.6  \\
    $b \bar b  (1, 10.7)$     &  51.9     &  41.8  &  10  & 0.5 \\
    $ t \bar t (1.1, 15.1) $   &  57  &   45.4  &   9.8  &   0.3  \\
    $ u \bar u (0.95, 6.5)$  &  49.4  & 40.3  &   10.7  &   1.0 \\ [1ex] 
\hline 
\end{tabular}
\label{table:table2} 
\end{table}

\begin{table}[!h]
\caption{ $ N_\gamma / {m^2_\chi} $ in units of $10^{-4} $ GeV$^{-2}$ for $m_\chi = 200 \, \textrm{GeV} $ } 
\centering 
\begin{tabular}{l c c c c } 
\hline\hline 
Channel & I  & II & III &  IV \\ [0.5ex] 
\hline 
    $ WW,ZZ (0.73, 7.76)$   &  13.5  &   11.0   &  3.1   &  0.4  \\
    $b \bar b  (1, 10.7)$     &  18.4     &  15.0  &  4.1  & 0.4 \\
    $ t \bar t (1.1, 15.1) $   &  20.2  &   16.4  &   4.3  &   0.3  \\
    $ u \bar u (0.95, 6.5)$  &  17.5  & 14.4  &   4.2  &   0.6 \\ [1ex] 
\hline 
\end{tabular}
\label{table:table3} 
\end{table}

\subsection{Estimating $< EM >$}

The geometry of dark matter caustics depends on the spatial dark matter velocity distribution. In the linear velocity field approximation~\cite{ref31}, the velocity field is a linear function of position, i.e. $\vec v(\vec r) = M \, \vec r$ where $M$ is a matrix. In general, $M$ contains both anti-symmetric and symmetric parts. If $M$ is dominated by the anti-symmetric part (rotational flow), the infall is greatly simplified and the resulting caustics have the appearance of rings~\cite{ref29}. If $M$ contains a significant symmetric part, the caustic geometry is more complicated. See~\cite{ref31} for a description. In this article, we assume that the caustics have the appearance of rings. If the flow has axial symmetry, the caustic ring is circular with constant cross section. Otherwise, the cross section will vary along the ring and the ring will not be circular. 

Let us consider cylindrical co-ordinates $(s,z)$ where $s = \sqrt{x^2 + y^2}$.  We assume axially symmetric infall about the $z$ axis and reflection symmetry about the $z=0$ plane. We can then obtain an analytic solution for the dark matter density at points close to the caustic. With these assumptions, the caustic is circular with a tricusp cross-section described by the curve
\begin{eqnarray}
z = 
  \left \{     
    \begin{array}{ll}
        \pm \frac{q}{2}\sqrt{ 1 - \frac{2}{3} \xi \pm \frac{8}{27} \xi^{3/2} - \frac{\xi^2}{27} }   &   \mbox { $0 \le \xi \le 1$ } \\
\\

        \pm \frac{q}{2}\sqrt{ 1 - \frac{2}{3} \xi + \frac{8}{27} \xi^{3/2} - \frac{\xi^2}{27} }   &   \mbox { $1 < \xi \le 9$  }
    \end{array}
\right.  \label{tricusp_curve}
\end{eqnarray}
where
\begin{eqnarray}
\xi &=& 1 + 8 \frac{s-a}{p}.  
\end{eqnarray}
Here, $a$ is the caustic radius ($s$ co-ordinate of the point of closest approach of the particles in the $z=0$ plane), $p$ and $q$ are the horizontal and vertical extents of the tricusp. See~\cite{ref29} for a detailed description.
 
Let us use the self-similar infall model~\cite{ref46,ref47} to estimate the mass infall rate for the flow of particles forming the caustic
\begin{equation}
\frac{d^2\,M}{d\Omega dt} = v\,f\,\frac{v^2_{rot}}{4\pi G}
\end{equation}
where $v$ is the velocity of the particles forming the flow, $f$ is a dimensionless quantity that characterizes the density of the flow and $v_{rot}$ is the rotation speed of the galaxy.

Let us define the two dimensionless co-ordinates $R = (s-a)/p$ and $Z = z/q$.  In the $Z=0$ plane, the density is given by
\begin{equation}
\rho(R,0) \approx 0.34 \frac{\textrm{GeV}}{\textrm{cm$^3$}} \; \frac{(f/10^{-2})\;( V_{rot} / 220 \, \textrm{km\,s$^{-1}$})^2 } {s_{kpc} \; p_{kpc}}   \left \{
    \begin{array}{lll}
         \frac{1}{1-R}   &    \mbox{ when $R \le 0$   } \\
         \frac{1}{1-R} \left( 1 + \frac{1}{\sqrt{R}} \right )  &  \mbox{ when $0 \le R \le 1$ } \\
         \frac{1}{R-1}   &    \mbox{ when $R \ge 1$   } 
     \end{array}
  \right.
\end{equation}
 $s_{kpc}$ and $p_{kpc}$ are distances measured in kpc.  For $R \rightarrow 0$ the dark matter density diverges as $R^{-{1/2}}$ for $ R > 0$. For $R \rightarrow 1$, the divergence $ \sim |(R-1)|^{-1}$.   When $Z \ne 0$, the density takes the form~\cite{ref48}
\begin{equation}
\rho(R,Z) \approx 0.17\, \frac{\textrm{Gev}}{\textrm{cm}^3} \; \frac{(f/10^{-2})\;( V_{rot} / 220 \, \textrm{km\,s$^{-1}$})^2 } {s_{kpc} \; p_{kpc}}  \sum_i \frac{1}{ \left| 2\,T_i \left( T_i - \frac{3}{2} \right ) + \left(1 - R\right ) \right | }
\end{equation}
where $T_i$ are the real roots of the quartic
\begin{equation}
T^4 - 2\,T^3 + (1-R)\,T^2 - \frac{27}{64} Z^2 = 0
\end{equation}
The above formulae are only valid at points close to the caustic.

The emission measure is calculated by integrating the density squared along the line of sight. Let $b = \pi/2 - \theta $ be the galactic latitude. $l$ is the galactic longitude chosen so that the galactic center is located in the direction $l=0, b=0$. We will assume that the caustics are spread over a distance $ \sim 10^{-4}$ pc. $f$ is set equal to $2 \times 10^{-2}$~\cite{ref14}. The cut-off density close to the fold surface(near $R=0, Z=0$) is then $ \approx 2.15 \times 10^3 / a \sqrt{p} $ Gev/cm$^3$. The density close to the cusp will be larger than this (the density falls off as the inverse of the distance to the cusp), but we will use the density close to the fold surface to set the density cut-off.

\begin{figure}[ht]
\subfigure[$\,(a=7.5,p=0.5,q=0.5)$ kpc]{\label{fig3-a}\includegraphics[width=3in,height=2in]
{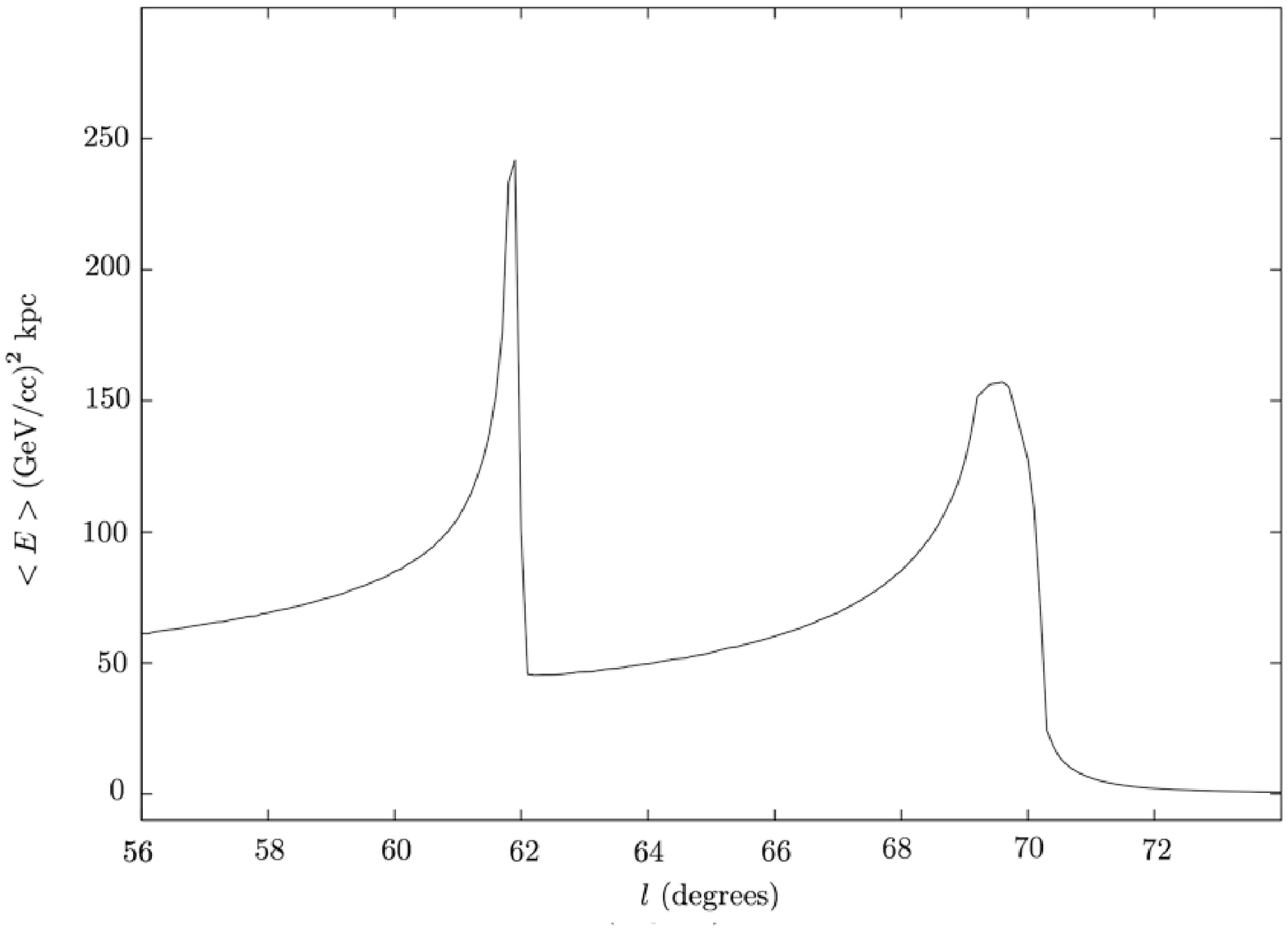}}
\subfigure[$\,(a=8.0,p=0.1,q=0.2)$ kpc]
{\label{fig3-b}\includegraphics[width=3in,height=2in]
{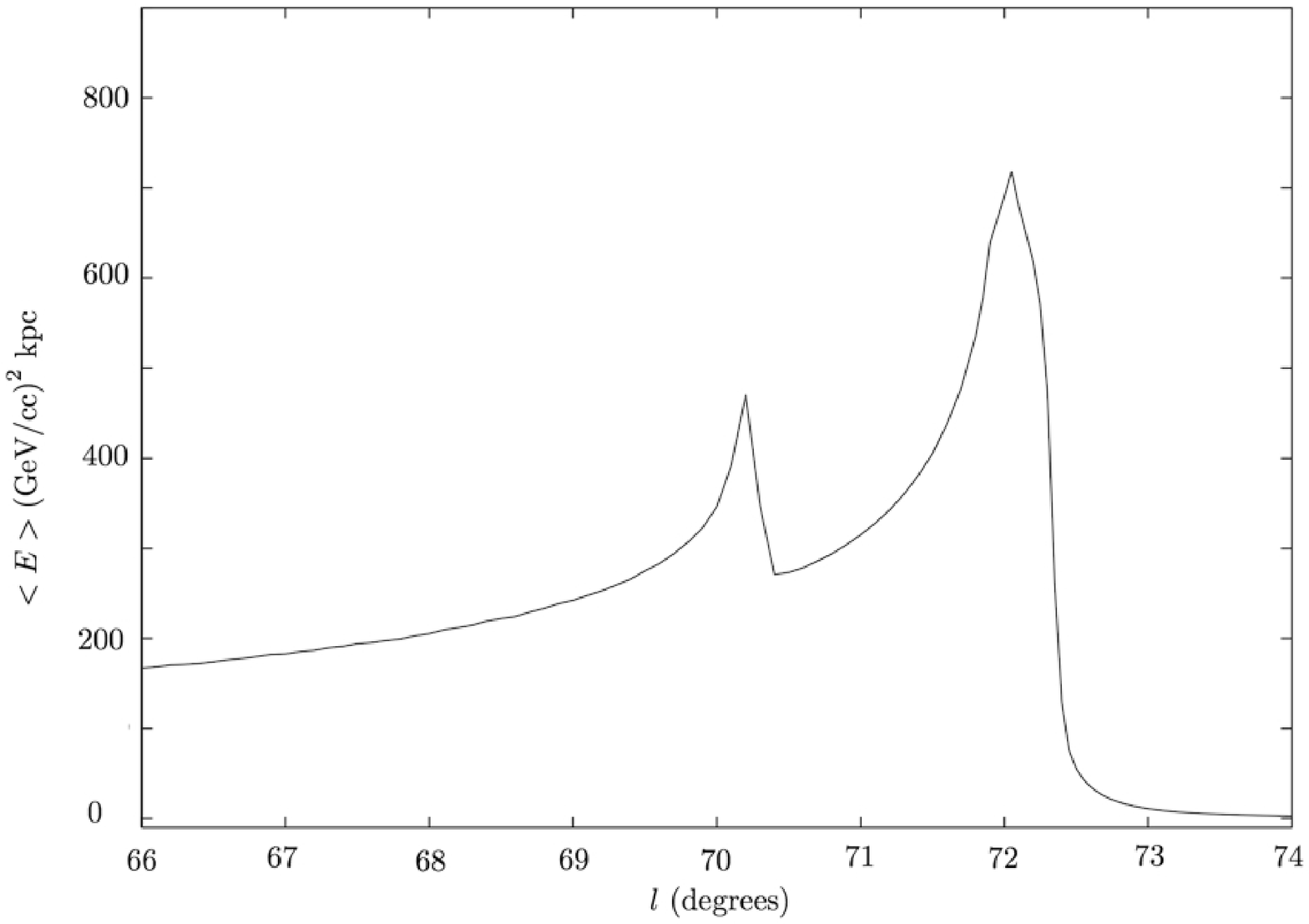}}
\subfigure[$\,(a=8.0,p=0.1,q=0.5)$ kpc]
{\label{fig3-c}\includegraphics[width=3in,height=2in]
{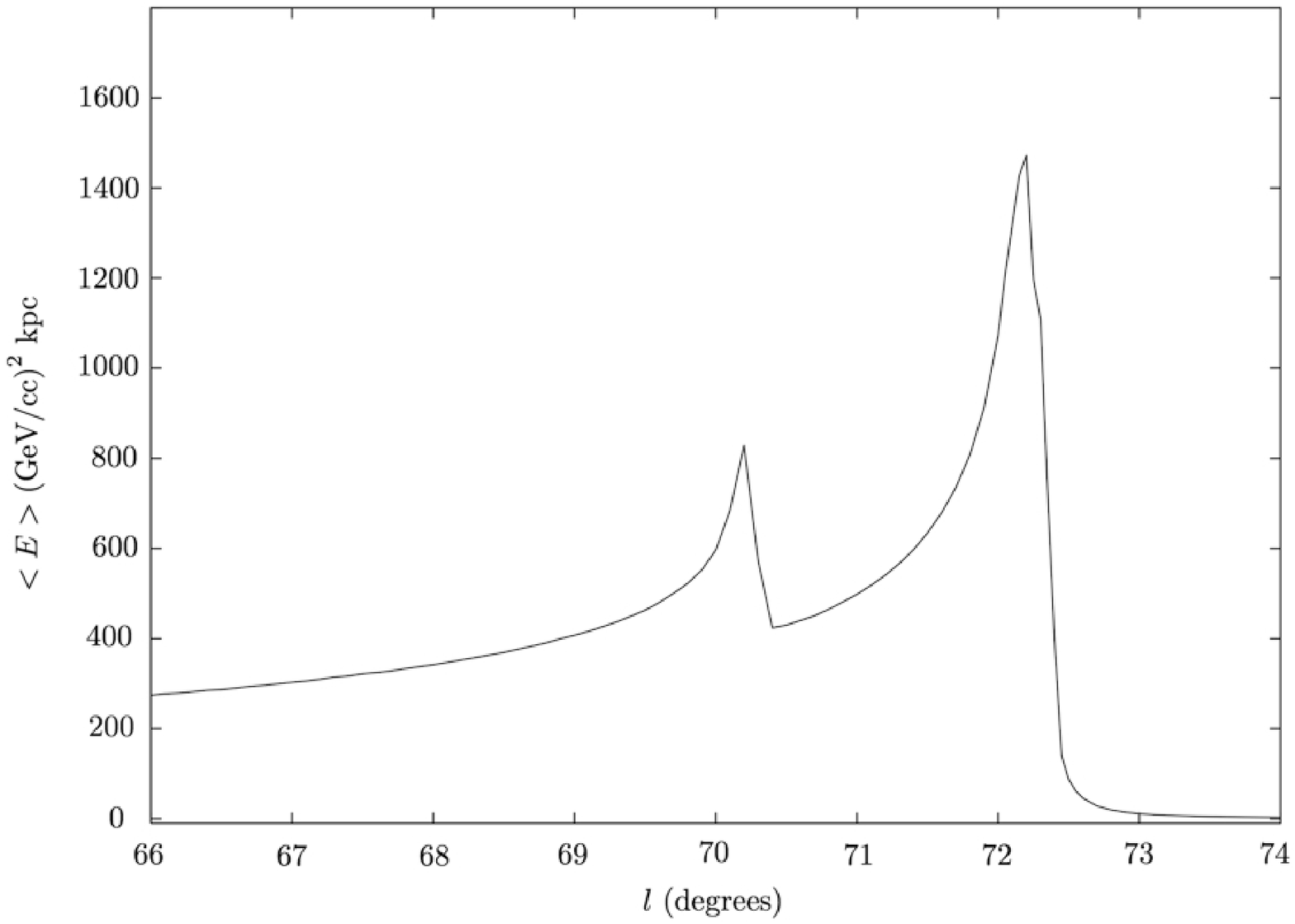}}
\caption{  Figures~\ref{fig3-a},~\ref{fig3-b},~\ref{fig3-c} show the emission measure averaged over a solid angle $\Delta\Omega = 10^{-5}$ sr for the three sets of caustic parameters.  
\label{fig3}}
\end{figure}

Fig.~\ref{fig3} shows the emission measure averaged over a solid angle $10^{-5}$ sr for three different sets of caustic parameters, as a function of longitude $l$. $b$ is set equal to 0 and we assume that the caustic lies in the galactic plane. Figures~\ref{fig3-a},~\ref{fig3-b} and~\ref{fig3-c} are plotted for $(a,p,q) = (7.5,0.5,0.5), (8.0,0.1,0.2)$ and $(8.0,0.1,0.5)$ respectively with all distances measured in kpc. The earth's location is set equal to $8.5$ kpc from the center. We expect the signal to be strongest when the line of sight is tangent to the ring. From the figures, we see that the emission measure is sensitive to the caustic geometry. The prominent features are the pair of peaks, or `hot spots' separated by a few degrees. The first peak occurs when the line of sight is tangent to the fold surface (when $s = a$). The second peak occurs when the line of sight is tangent to the cusp line (when $s = a+p$). (In the limit $p,q \rightarrow 0$, the two peaks coincide, see~\cite{ref12}). For the case when $p=0.5$ kpc (Fig.~\ref{fig3-a}), the cut-off density was set equal to $400$ GeV/cm$^3$ everywhere, while for the plots with $p = 0.1$ kpc (Figs~\ref{fig3-b} and~\ref{fig3-c}), the cut-off density was set equal to $800$ GeV/cm$^3$ everywhere. The magnitudes of $< EM >$ for the hot spots depend on the values of the caustic parameters and also on the averaging scale (here chosen to be $10^{-5}$ sr).  Table IV shows the annihilation flux for the two hot spots for different values of the averaging scale $\Delta\Omega$, for the case $(a=8.0, p=0.1, q=0.2)$ kpc. It is worth pointing out that if the triangular feauture in the IRAS map is interpreted as the imprint of the nearest caustic on the surrounding gas as in~\cite{ref40}, the implied caustic parameters are close to what we have assumed for Fig~\ref{fig3-b}.

\begin{table}[!h]
\caption{ Peaks of $< EM >$ for $(a=8.0, p=0.1, q=0.2)$ kpc} 
\centering 
\begin{tabular}{ | c | c | c | } 
\hline\hline 
$\Delta\Omega$   &   Fold peak  &  Cusp peak \\
(sr)  &  (GeV/cc)$^2$ kpc  &  (GeV/cc)$^2$ kpc   \\ [1ex]  

\hline 

$10^{-7}$ &   1549.8  &   2177.6 \\
$10^{-6}$ &   1048.3  &   1283.3 \\ 
$10^{-5}$ &    469.9 &   718.1 \\
$10^{-4}$ &   269.6 &    239.4 \\
$10^{-3}$ &   115.4   &   52.0 \\ [1ex] 
\hline 
\end{tabular}
\label{table:table4} 
\end{table}

\section{Comparing the signal with the background}

The annihilation flux from caustics is thus given by
\begin{eqnarray}
\Phi &=& S \times \frac{< EM>}{4\pi} \nonumber \\
       &\approx & 110 \frac{N_\gamma}{ \left( m_\chi/100 \; \textrm{GeV} \right )^2 } \; \frac{ < EM > }{ (\textrm{Gev/cc})^2 \; \textrm{kpc} } \; \left[ \textrm{ m$^2$ sr year }\right]^{-1} 
\label{flux}
\end{eqnarray}
Let us compare this flux with the expected background. The EGRET measured background flux $\Phi_{bg}(E_1,E_2)$ from energies $E_1$ to $E_2$ is given by~\cite{hunter,ref44}
\begin{equation}
 \Phi_{bg} (E_1,E_2,\theta,\phi) = N_{\gamma,bg}(E_1,E_2) \times N_0(\theta,\phi) \times 10^{-6} \textrm{ cm$^{-2}$ s$^{-1}$ sr$^{-1}$ }
\end{equation}
where 
\begin{equation}
N_{\gamma,bg}(E_1,E_2) = \int_{E_1}^{E_2} dE\, \left(\frac{E}{\textrm{GeV}}\right)^{-2.7}
\end{equation}
The function $N_0(\theta,\phi)$ is energy independent and follows the fitting form given in~\cite{ref44}. For the four energy bands we have considered, $N_{\gamma,bg}$ = 198.8 for Band I (30 MeV - 100 MeV),  28.9 for Band II (100 MeV - 1 GeV), 0.58 for Band III (1 GeV - 10 GeV) and     0.012 for Band IV (above 10 GeV).

We now compare the caustic signal with the expected background. Since the gamma ray background falls off with energy ($E^{-2.7}$) faster than the annihilation signal ($E^{-1.5}$), we expect that the best chance for detection is at moderately high energies. At low energies, the background flux overwhelms the signal, while at very high energies, the signal is weak. We choose $m_\chi = 50$ GeV since this choice gives the largest flux. For the quantity $ N_\gamma / {m^2_\chi} $, we use the average value for the band. The averaging scale $\Delta\Omega$ is set to $10^{-5}$ sr.

\begin{figure}[ht]
\subfigure[$\,$Energy Band III]{\label{fig4-a}\includegraphics[width=3in,height=2in]
{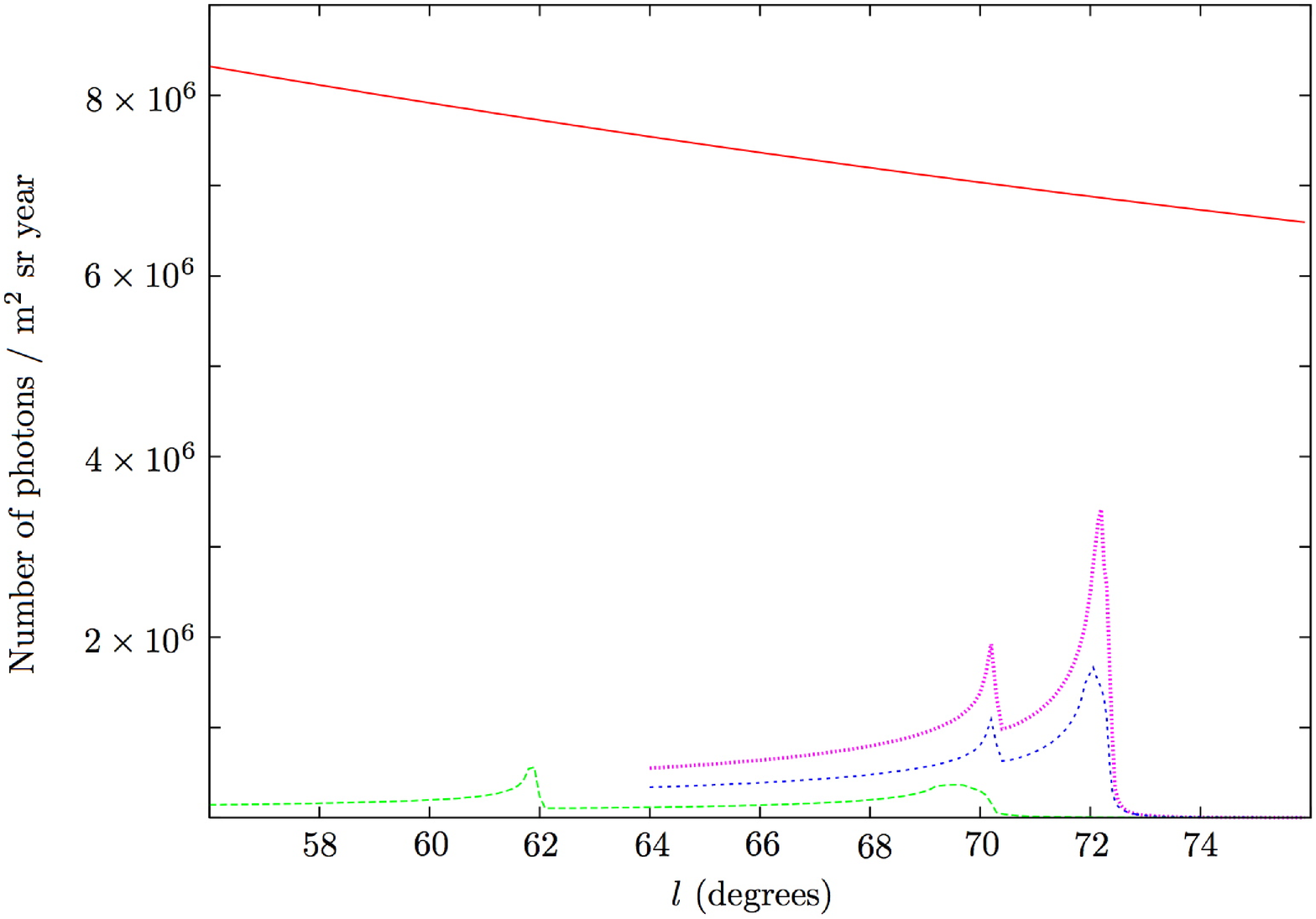}}
\subfigure[$\,$Energy Band IV]
{\label{fig4-b}\includegraphics[width=3in,height=2in]
{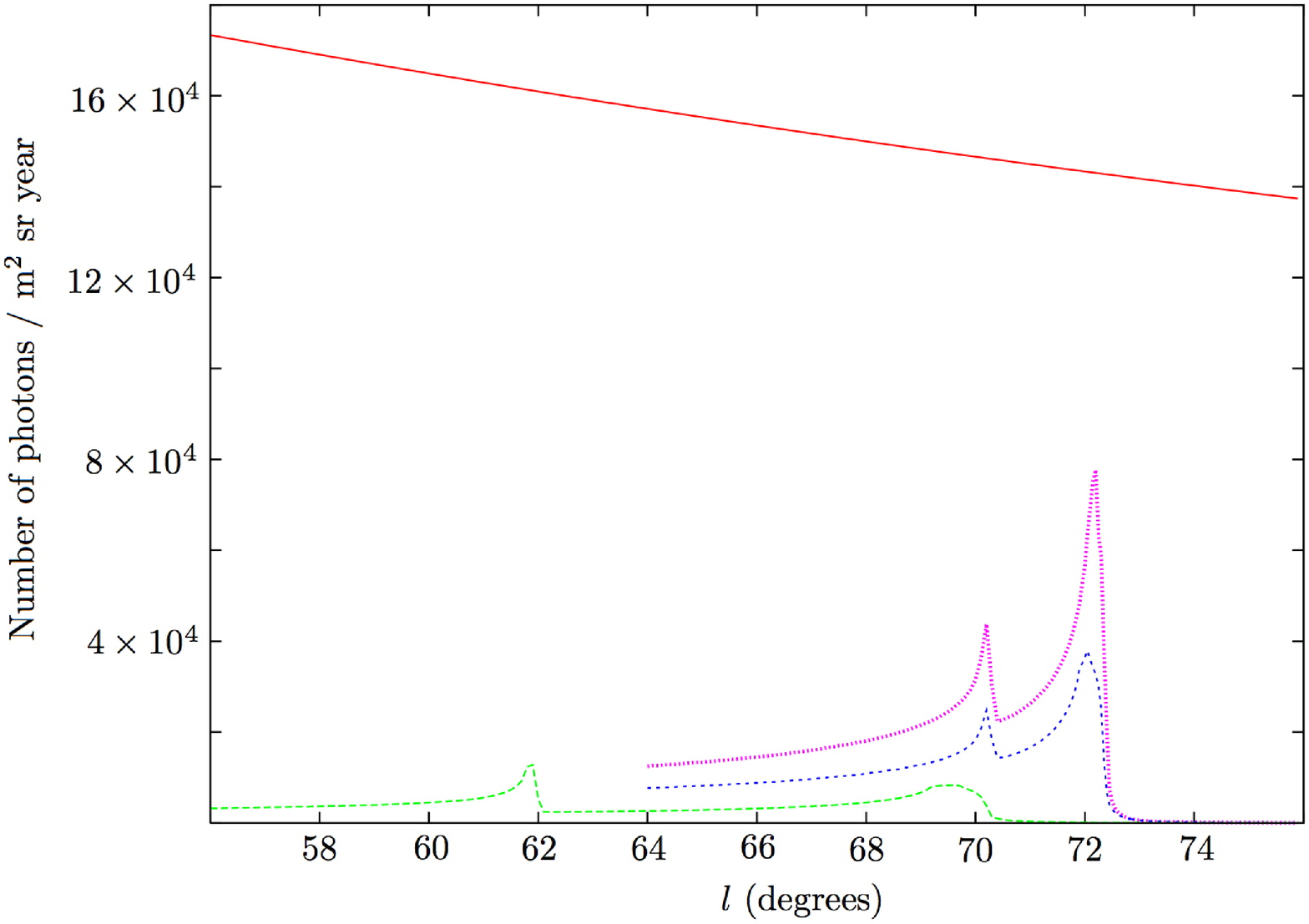}}
\caption{  (color online) The annihilation flux for the three different sets of caustic parameters ($m_\chi = 50$ GeV), compared with the EGRET measured diffuse background. The background signal appears as a line of nearly constant slope. The characteristic peaks and the sharp fall-off of the signal may help in signal detection. Among the three curves that show the signal flux, the largest flux curve corresponds to the case $(a=8.0,p=0.1,q=0.5)$, the middle curve is for $(a=8.0,p=0.1,q=0.2)$ and the smallest flux is for $(a=7.5,p=0.5,q=0.5)$. Energy Band III is favored to Energy Band IV owing to the very small flux in Band IV. 
\label{fig4}}
\end{figure}

Figures~\ref{fig4-a} and~\ref{fig4-b} show the expected annihilation flux (number of photons per square meter, per steradian, per year) as a function of angle $l$ near the plane of the galaxy ($b = 0$) for the three sets of caustic parameters we considered, for Energy Bands III and IV respectively. This is contrasted with the expected diffuse gamma ray background. The figure shows only the flux from the caustic and does not include the smooth component of the halo. For a standard isothermal with core type of halo profile, we expect the signal from the smooth component of the halo to be much smaller than the signal from the caustic. However,  if there are dark matter clumps along the line of sight, annihilation from the clumps may be significant \cite{ms}. In principle, the peaks in the signal and the sharp fall-off of flux are helpful in identifying the annihilation signal, particularly for the more optimistic caustic parameters and for small WIMP masses. For large WIMP masses, the annihilation signal is significantly smaller.

Let us estimate the significance of detection by assuming an angular resolution $\Delta\Omega = 10^{-5}$ sr, an integration time of 1 year, a detector area of 1 $m^2$ and an optimistic value of $m_\chi = 50$ GeV. Further let us consider the 1 GeV - 10 GeV band, for which the mean value of $ \frac{N_\gamma}{(m_\chi / 100 \, GeV)^2}$ is 21.1 (Table \ref{table:table1}). The number of photons received, assuming $100\%$ detector efficiency, from a region with angular extent $\Delta\Omega = 10^{-5}$ sr is then (Eq. \ref{flux}), $ N_{\gamma,signal} \approx 0.02 \, \frac{< EM >}{(GeV/cc)^2 \, kpc}$. Since an angular size of $10^{-5}$ sr corresponds to a spot of diameter roughly $0.2^\circ$ at the equator, we may expect about 10 distinguishable spots in the angular range $-0.1^\circ < b < 0.1^\circ, 68.2^\circ < l < 70^\circ$. Here $b$ is the latitude ($\pi/2 - \textrm{polar angle} \; \theta$) and $l$ is the longitude. For the moderately optimistic case of caustic parameters $(a=8,p=0.1,q=0.2)$ kpc, we find approximately 60 signal events and for the very optimistic case $(a=8,p=0.1,q=0.5)$ kpc, we find approximately 101 signal events. In this range, we find 716 background events.  Assuming that the fluctuation in the background varies as the square root of the number of background events ($N_{\gamma,bg}$), we may define the detection significance $\sigma = N_{\gamma,signal} / \sqrt{N_{\gamma,bg}}$. For the case $(a=8,p=0.1,q=0.2)$ kpc, we find $\sigma = 2.2$ while for $(a=8,p=0.1,q=0.5)$ kpc, we find $\sigma = 3.8$. A similar calculation shows that in the range $-0.1^\circ < b < 0.1^\circ, 70.2^\circ < l < 72^\circ$, we find 93 signal events for the case $(a=8,p=0.1,q=0.2)$ kpc, and 148 signal events for the case $(a=8,p=0.1,q=0.5)$ kpc, with 700 background events. The values of $\sigma$ for the two sets of caustic parameters is found to be 3.5 and 5.6 respectively.

\section{Conclusions}

We calculated the gamma ray annihilation signal from a nearby dark matter caustic having the geometry of a ring with a tricusp cross section near the plane of the galaxy, in different energy bands. For such a caustic, the annihilation signal has two peaks, separated by a few degrees, depending on the size of the caustic. There is an abrupt fall-off of flux after the second peak.  Since the diffuse gamma ray background flux falls off with energy faster than the signal, it is advantageous to look for the signal at moderately high energies. We compared the expected annihilation flux with the expected diffuse gamma ray background. The characteristics of the annihilation flux can in principle, be used to discriminate between the signal and the background. In practice however, we expect this to be a challenging task.

\acknowledgments{ I thank Paolo Gondolo, Konstantin Matchev, Pierre Sikivie and Richard Woodard for helpful discussions. This work was supported by the U.S. Department of Energy under contract DE-FG02-97ER41029. }

\end{document}